\documentclass[aip,reprint]{revtex4-1}

\usepackage[separate-uncertainty = true]{siunitx}
\usepackage[format=plain,justification=Justified, textfont=small]{caption} 
\usepackage{subcaption}
\usepackage{amsmath,amssymb}
\DeclareSIUnit\sq{\ensuremath{\Box}}
\usepackage{graphicx}

\draft 

\begin{document}



\title{A 64-pixel mid-infrared single-photon imager based on superconducting nanowire detectors} 



\author{Benedikt Hampel}
\email[]{benedikt.hampel@nist.gov}
\affiliation{National Institute of Standards and Technology, Boulder, Colorado 80305, USA}
\affiliation{Department of Physics, University of Colorado, Boulder, Colorado 80309, USA}

\author{Richard P. Mirin}
\author{Sae Woo Nam}
\author{Varun B. Verma}
\affiliation{National Institute of Standards and Technology, Boulder, Colorado 80305, USA}


\date{28 September 2023}
\begin{abstract}
     A large-format mid-infrared single-photon imager with very low dark count rates would enable a broad range of applications in fields like astronomy and chemistry. Superconducting nanowire single-photon detectors (SNSPDs) are a mature photon-counting technology as demonstrated by their figures of merit. However, scaling SNSPDs to large array sizes for mid-infrared applications requires sophisticated readout architectures in addition to superconducting materials development. In this work, an SNSPD array design that combines a thermally coupled row-column multiplexing architecture with a thermally coupled time-of-flight transmission line was developed for mid-infrared applications. The design requires only six cables and can be scaled to larger array sizes. The demonstration of a 64-pixel array shows promising results for wavelengths between \SI{3.4}{\micro\meter} and \SI{10}{\micro\meter}, which will enable the use of this single-photon detector technology for a broad range of new applications.
\end{abstract}


\maketitle 

Superconducting nanowire single-photon detectors (SNSPDs) have matured into a widespread photon-counting technology for applications in fields like quantum optics,\cite{Lita.2022} quantum computing,\cite{Hampel.2023} and quantum communications \cite{Gruenfelder.2023}. Some of their outstanding metrics are very high system detection efficiencies of over \SI{98}{\percent},\cite{Reddy.2020} low timing jitter below \SI{3}{\pico\second},\cite{Korzh.2020} dark count rates below \num{1}~count per day,\cite{Chiles.2022} high maximum count rates up to \SI{1.5}{gigahertz},\cite{Craiciu.2023} zero readout noise, and sensitivity over a broad range of wavelengths from the ultraviolet (UV) \cite{Wollman.2017} to the mid-infrared \cite{Verma.2021}.  

Cameras capable of counting single mid-infrared photons could be an important technological development for applications in astronomy, in particular for the spectroscopy of exoplanet atmospheres in future space telescopes.\cite{Wollman.2021} Mid-infrared cameras used in current space telescopes such as the James Webb Space Telescope utilize blocked impurity band (BIB) and mercury cadmium telluride (MCT) detectors. Although they operate at higher temperatures than SNSPDs (\SIrange[range-phrase= \text{ -- } ]{4}{10}{\kelvin}), these detectors do not have sufficient sensitivity to count single photons. BIB detectors also suffer from drawbacks such as reset anomalies, last-frame effects, droop, and drift which all lead to instability, making them less desirable for applications such as exoplanet spectroscopy which requires stability on the order of only a few ppm over several hours.\cite{Ressler.2015, Greene.2016} Both BIB and MCT detectors also exhibit significantly higher noise and dark count rates in comparison to SNSPDs. In addition to applications in astronomy, mid-infrared-optimized SNSPDs have shown promise in the field of physical chemistry and vibrational spectroscopy.\cite{Lau.2023} 

While there are several potential applications for mid-infrared-optimized single-photon detector arrays, in the recent past SNSPDs have only been used at telecommunications and visible wavelengths. The optimization of SNSPDs for mid-infrared wavelengths is challenging due to the materials development needed to achieve high detection efficiencies of photons with lower energies. Both the superconducting material used to create the SNSPD as well as the materials used in the optical stack which enhances the absorption in the SNSPD must be modified compared to telecom and shorter wavelengths. Recently, we demonstrated that SNSPDs based on the amorphous superconductor WSi are capable of detecting single photons up to a wavelength of \SI{10}{\micro\meter}.\cite{Verma.2021, Verma.2022} This was achieved by increasing the silicon content of the WSi resulting in a lower superconducting gap energy and higher resistivity. 

In addition to the challenges inherent in improving the long-wavelength sensitivity of single pixels, large-format SNSPD arrays are still at an early stage of development even in the near-infrared. Various multiplexing techniques have been demonstrated over the past decade including row-column readout,\cite{Allman.2015, Wollman.2019} thermal row-column,\cite{Allmaras.2020} and thermally-coupled imager (TCI) \cite{McCaughan.2022, Oripov.2023}. To date, the largest array of 400 kilopixels was demonstrated using the TCI architecture,\cite{Oripov.2023} although this array was only efficient at detecting photons in the visible.

The high-resistivity WSi films and narrow nanowire widths used in the fabrication of mid-infrared-optimized SNSPDs result in very low bias currents on the order of \SI{1}{\micro\ampere}.\cite{Verma.2021} This makes multiplexing schemes such as TCI more challenging, primarily because this scheme relies on thermal coupling between the SNSPDs and a superconducting transmission line bus while the amount of heating scales quadratically with the SNSPD bias current. With typical near-infrared SNSPDs having bias currents on the order of \SI{10}{\micro\ampere}, mid-infrared-optimized SNSPDs will generate only \SI{1}{\percent} the amount of Joule heating compared to near-infrared SNSPDs when a photon is detected, which may not be enough to thermally trigger the bus. Thus, the TCI multiplexing scheme must be significantly modified to develop SNSPD arrays that can be used as mid-infrared cameras in telescopes for astronomy or physical chemistry.

\begin{figure*}
    \centering
    \includegraphics[width=1.00\textwidth]{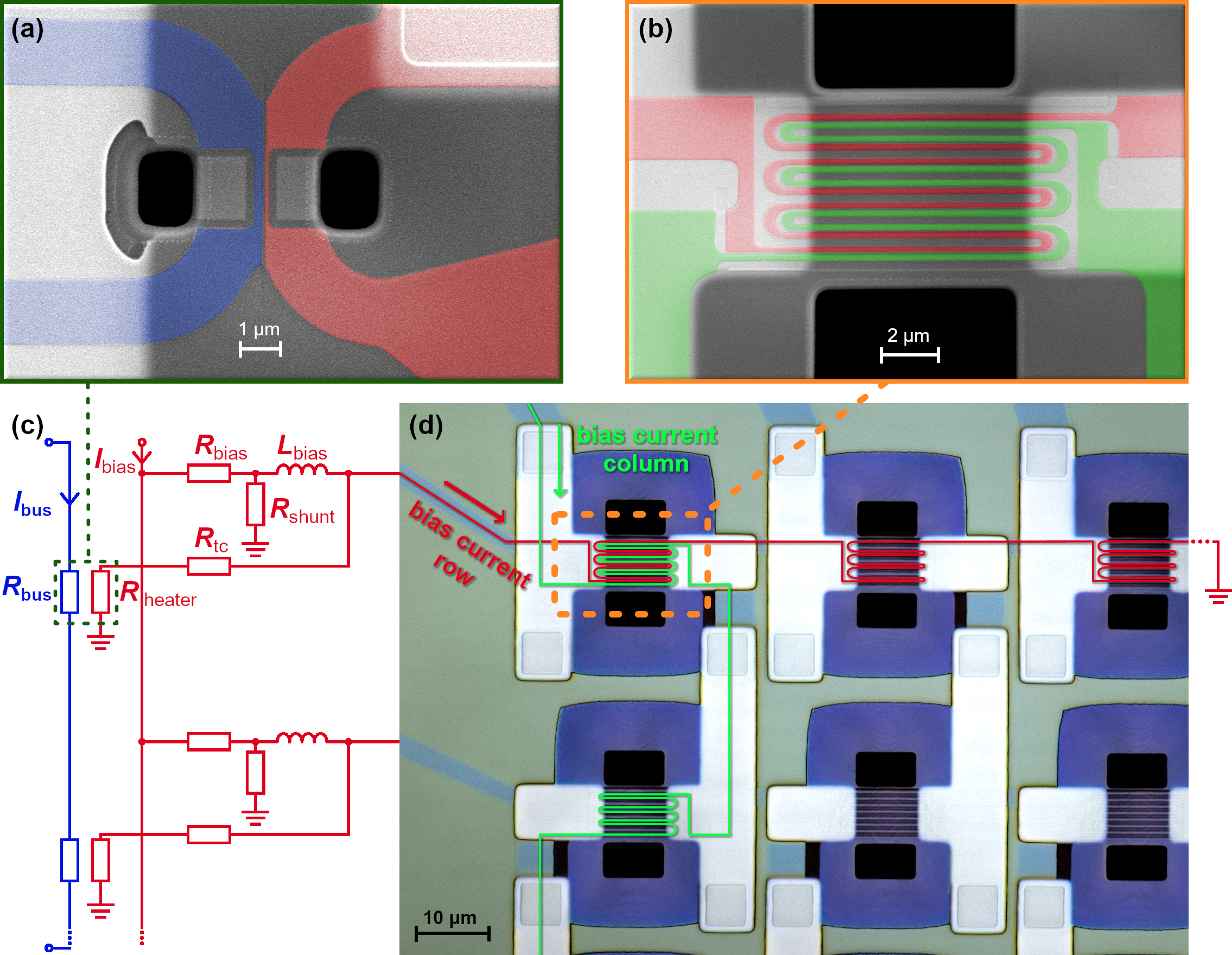}
    \vspace{-3mm}
    \caption{Micrographs of a 64-pixel mid-infrared-optimized SNSPD array. (a) Scanning electron micrograph (SEM) of the thermal coupling from the SNSPDs (false colored in red) to the transmission line (false colored in blue). (b) SEM of a single pixel with interleaved row (false colored in red) and column (false colored in green) SNSPDs. (c) Circuit diagram of the bias current network that connects to the SNSPD array (red) and is thermally coupled to the transmission line (blue). (d) Photomicrograph of six SNSPD pixels with highlighted current paths for one column (green) and one row (red). }
    \label{Overview}
\end{figure*}

The SNSPD array that is presented in this work was designed combining the thermally-coupled row-column multiplexing architecture \cite{Allmaras.2020} with a thermally-coupled imager architecture.\cite{McCaughan.2022, Oripov.2023} The details of the design are depicted in Fig.~\ref{Overview}. Each pixel consists of two interleaved SNSPDs, which are false colored in red and green in the scanning electron micrograph (SEM) in Fig.~\ref{Overview}(b) highlighting the co-wound row and column SNSPD, respectively. All SNSPDs in a single row (red) or column (green) are wired in series as shown in Fig.~\ref{Overview}(d). All rows are current-biased in parallel supplied through a single cable as depicted in red in Fig.~\ref{Overview}(c). The current \textit{I}\textsubscript{bias} is divided into all the rows of the full array and biases the row SNSPDs through the bias resistors \textit{R}\textsubscript{bias} and the large bias inductors \textit{L}\textsubscript{bias}. Note that only the row bias and readout is shown in Fig.~\ref{Overview} for clarity. This bias and readout network is also present for the columns but is not shown. 

When a photon is absorbed in one of the co-wound nanowires (either row (red) or column (green)), it generates a resistive hotspot. The Joule heating resulting from the bias current flowing through the resistive domain of the nanowire generates phonons which are absorbed by the neighboring interleaved nanowire, resulting in the formation of a hotspot in that nanowire as well. Thus, a ``click'' in a row (or column) nanowire always results in a corresponding ``click'' in the corresponding column (or row) nanowire at that particular pixel. The SNSPDs are patterned on silicon dioxide (SiO\textsubscript{2}) membranes to enhance this thermal coupling. The breakdown of superconductivity in the SNSPDs results in a sudden change of resistance for the bias current which happens over a timescale of several picoseconds. The resulting current pulse is redirected out of the SNSPD into the circuit shown in red in Fig.~\ref{Overview}(c). Since the impedance of the bias inductor L\textsubscript{bias} is very large over the timescale of the pulse, the current pulse is primarily driven into the  path of the thermal coupling resistance \textit{R}\textsubscript{tc} and a heater resistance \textit{R}\textsubscript{heater}. The heater resistance is a constriction in a superconducting strip, which can be seen false colored in red in the SEM micrograph in Fig.~\ref{Overview}(a). The redirected bias current will exceed the critical current density in this constriction, superconductivity will break down, and the finite resistance will cause a localized heating in the constriction area. The phonons generated by the Joule heating are then relayed to an adjacent constriction in the superconducting transmission line bus seen false colored in blue in the SEM micrograph in Fig.~\ref{Overview}(a). The large series inductance \textit{L}\textsubscript{bias} ensures that sufficient energy is dissipated in the constriction to trigger a detection event on the transmission line bus, as this energy scales linearly with \textit{L}\textsubscript{bias} and quadratically with the bias current. The thermal coupling resistance \textit{R}\textsubscript{tc} sets the electrical recovery time of the bias current back into the SNSPDs after a detection event and is used to prevent latching of the SNSPDs. Finally, the shunt resistor \textit{R}\textsubscript{shunt} provides a path to ground for any leakage current through the large inductor \textit{L}\textsubscript{bias}. This helps to mitigate redistribution of bias current to other rows or columns on the same bias line during a detection event.

The readout busses for the row and column SNSPDs are designed as superconducting microstrip transmission lines and are current-biased with \textit{I}\textsubscript{bus}. Their impedance is transformed to match the cable impedance of \SI{50}{\ohm} by using Hecken tapers \cite{Hecken.1972} at each end of the line. The sudden increase in resistance which occurs when a constriction in the transmission line switches to the normal state results in a positive voltage pulse traveling to one end of the transmission line and a negative voltage pulse traveling to the other. These voltage pulses can be amplified and subsequently measured with a time tagger. The time difference between the positive and negative voltage pulse is finally used to determine which row or column had a detection event. The combination of the results for the row and the column transmission lines allows one to infer which pixel absorbed a photon. This design enables readout of the whole array with only six cables independent of array size. Two cables are used to bias the row and column SNSPDs, two cables for the row transmission line, and two cables for the column transmission line. 

A 64-pixel SNSPD array optimized for the mid-infrared was fabricated in the NIST Boulder Microfabrication Facility using the design described above. Further fabrication details are available in the Supplementary Material. 

\begin{figure*}
    \centering
    \begin{subfigure}[b]{0.47\textwidth}
        \centering
        \includegraphics[width=1.00\textwidth]{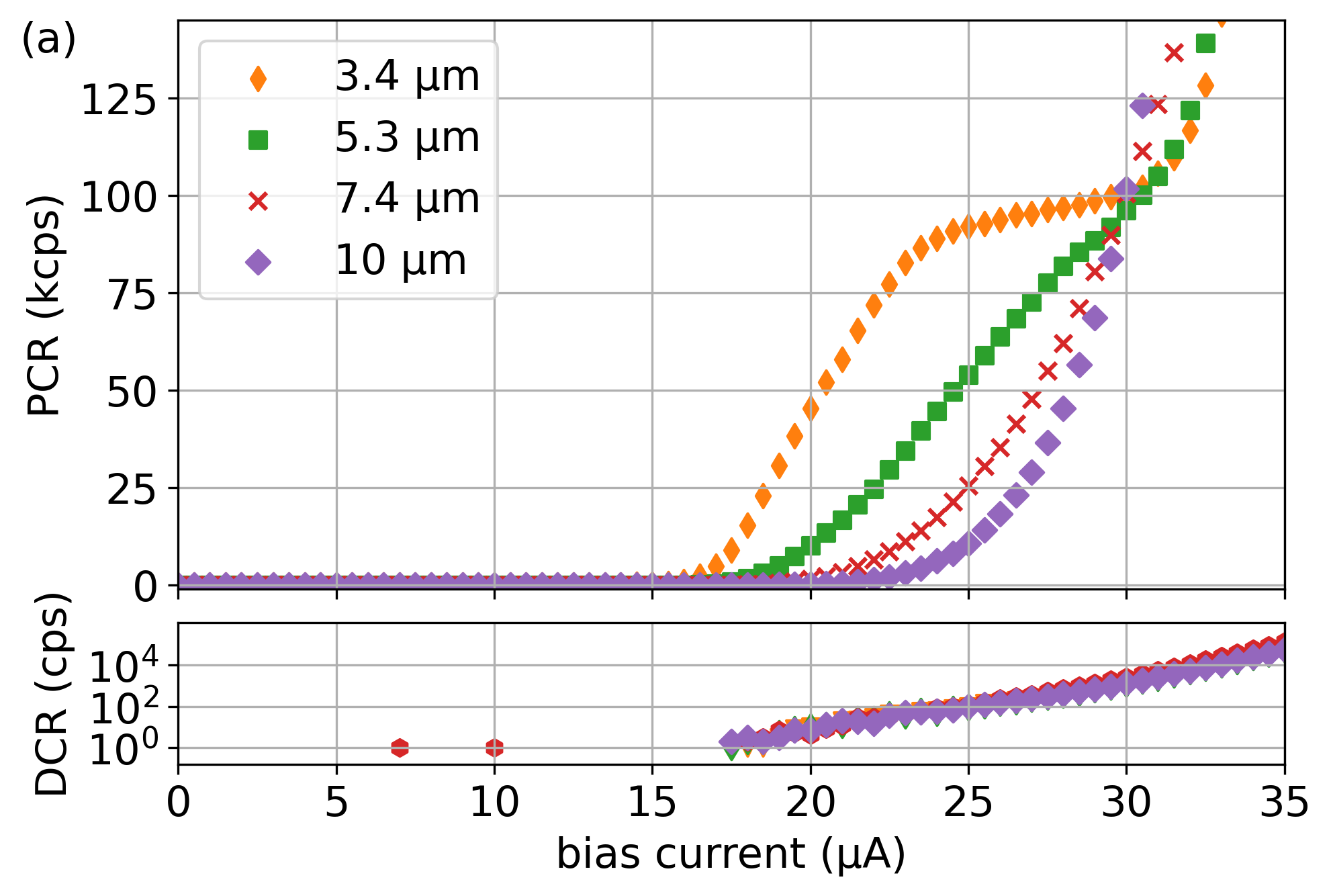}
        \label{BCR:rows}
        \vspace{-5mm}
    \end{subfigure}  
    \begin{subfigure}[b]{0.47\textwidth}
        \centering
        \includegraphics[width=1.00\textwidth]{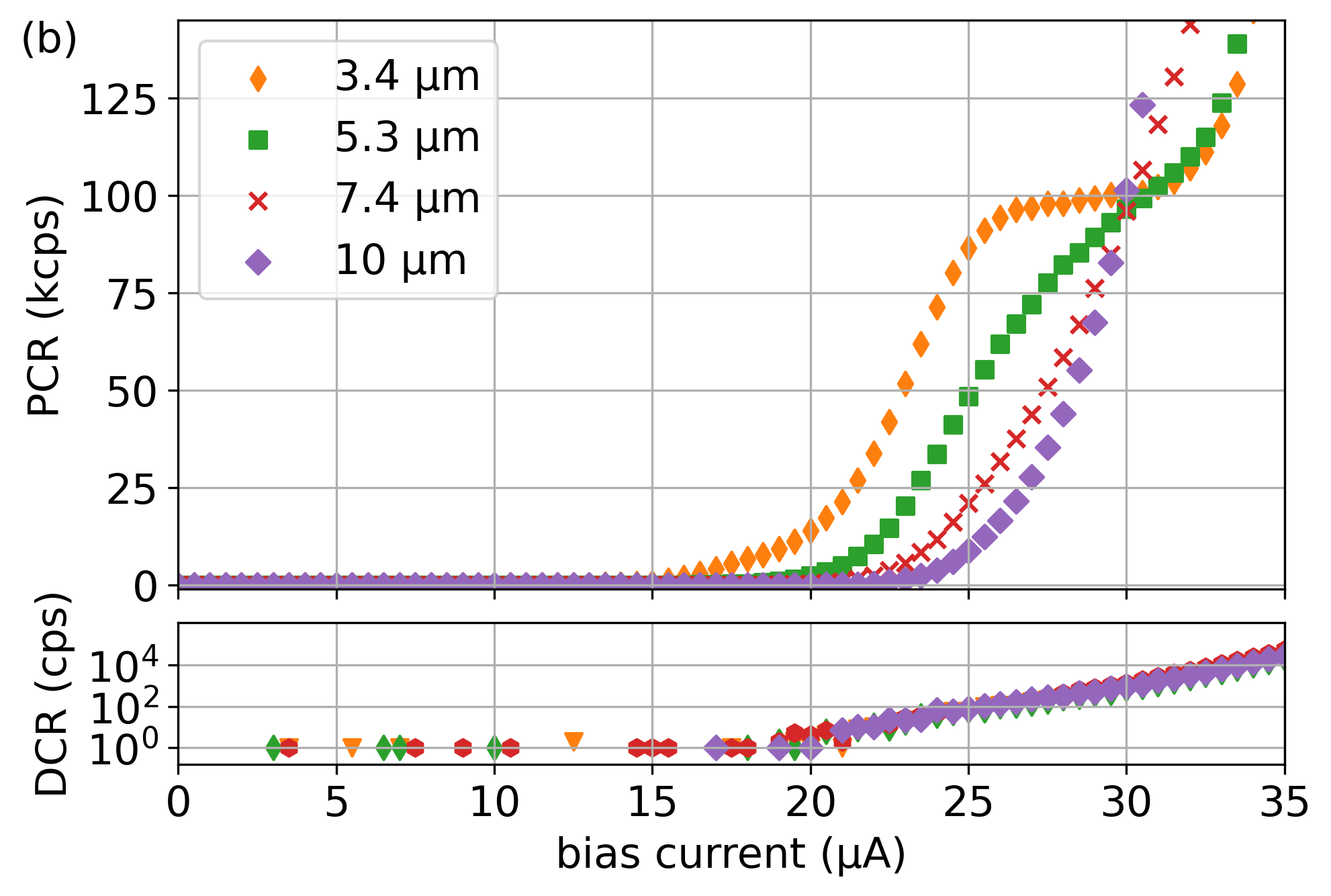}
        \label{BCR:cols}
        \vspace{-5mm}
    \end{subfigure} 
    \begin{subfigure}[b]{0.47\textwidth}
        \centering
        \includegraphics[width=1.00\textwidth]{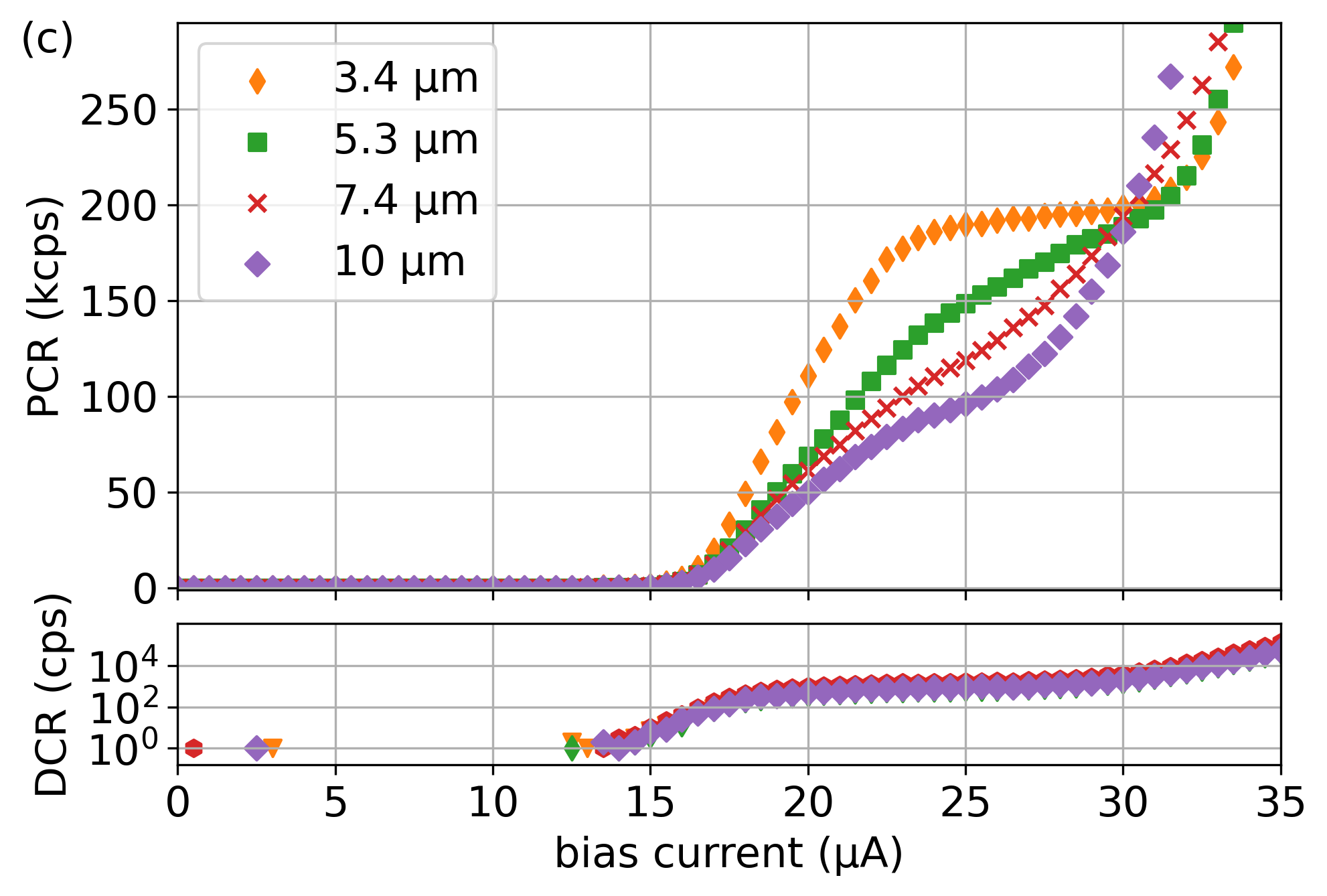}
        \label{BCR:rowsColon}
        \vspace{-5mm}
    \end{subfigure} 
    \begin{subfigure}[b]{0.47\textwidth}
        \centering
        \includegraphics[width=1.00\textwidth]{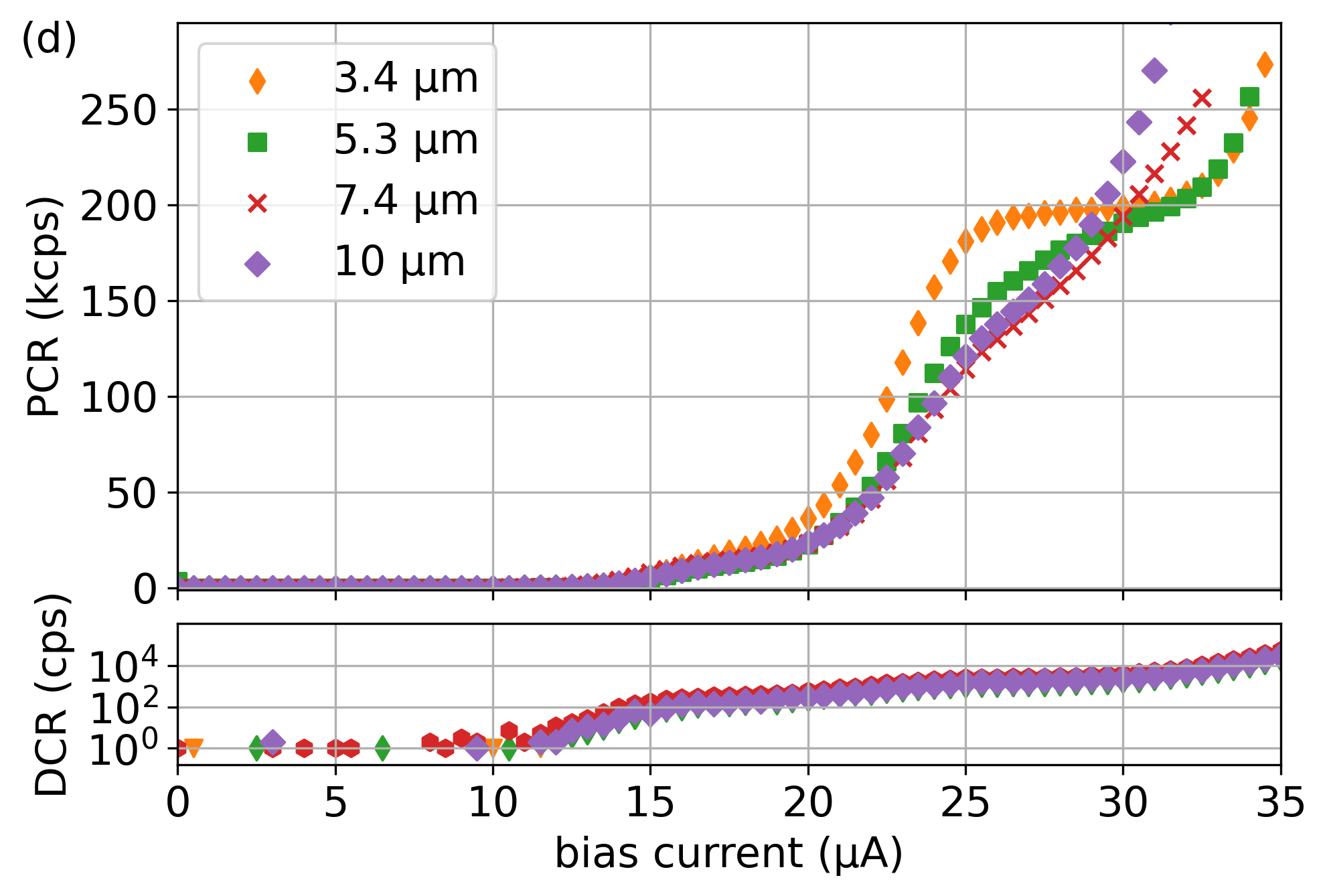}
        \label{BCR:colsRowson}
        \vspace{-5mm}
    \end{subfigure} 
    \caption{Photon count rate (PCR) and dark count rate (DCR) in counts per second (cps). Panels (a) and (c) show measurements on the row transmission line with (a) row bias varied, columns unbiased and (c) row bias varied, columns biased at \SI{30}{\micro\ampere}. Panels (b) and (d) show measurements on the column transmission line with (b) column bias varied, rows unbiased and (d) column bias varied, rows biased at \SI{30}{\micro\ampere}.}
    \label{BCR}
\end{figure*}

The cryogenic measurements were performed in an adiabatic demagnetization refrigerator (ADR) at a temperature of \SI{350}{\milli\kelvin}. The connection to the SNSPD bias network and the transmission lines was made by coaxial cables, which were connected to six bias-tees at room temperature outside the cryostat. The row and column SNSPDs were biased separately through the dc ports of two bias-tees. Each of the transmission lines was also biased through the dc port of one of the connected bias-tees, while the dc port of the bias-tee at the other side of the transmission line was terminated with a \SI{50}{\ohm} resistor. The ac ports of the transmission line bias-tees were amplified with two amplifiers with a total gain of \SI{44}{\decibel} and can be connected to either a counter or a time tagger. 

The 64-pixel array was flood-illuminated with a broadband thermal light source, which was filtered by an exchangeable narrow bandpass mid-infrared filter. A calcium fluoride (CaF\textsubscript{2}) window acted as a \SI{10}{\micro\meter}-lowpass filter to further attenuate any long wavelength blackbody radiation that is not perfectly blocked by the narrowband filter. The thermal light source was mounted at the \SI{4}{\kelvin}-stage of the cryostat. An additional copper aperture with a diameter of approximately \SI{3}{\milli\meter} was mounted in front of the filter stack and thermal source without making direct contact to the thermal source assembly. This shielding was also mounted at the \SI{4}{\kelvin}-stage and further reduced unwanted blackbody radiation from the thermal source irradiating the array. The array chip was mounted in a gold-coated copper housing with another aperture of \SI{2.5}{\milli\meter} diameter. There, the chip was bonded to connectors for the coaxial cables. 

The characterization of the photon count rate (PCR) as a function of bias current through the SNSPDs for different narrow bandpass filters is depicted in Fig.~\ref{BCR}. The transmission lines were biased with a current of \SI{2.8}{\micro\ampere} for all measurements as the count rates on both transmission lines were in a saturated regime at this value when the rows and columns were also biased. The bias currents for the SNSPDs were varied for different measurements. The thermal light source was pulsed with a function generator at a repetition rate of \SI{1}{\kilo\hertz} and duty cycle of \SI{50}{\percent}. Its power was adjusted for each of the narrow bandpass filters so that a PCR of \SI{100}{\kilo cps} (counts per second) could be measured for the row SNSPDs biased at \SI{30}{\micro\ampere}, while the column SNSPDs were not biased. All mentioned bias currents must be divided by a factor of \num{8} to obtain the bias current through a single SNSPD as all row and all column SNSPDs were biased in parallel through two bias cables as explained above. 

Each measurement started with a characterization of the dark count rate (DCR). For this, the thermal light source was switched off, the bias current was ramped up, and the count rates were measured with a counter at one end of the corresponding transmission line for \SI{1}{\second} per bias current value. The results are shown under each of the measurements on a logarithmic scale in Fig.~\ref{BCR}. A large part of the observed DCR can be attributed to unwanted blackbody radiation of the measurement setup. The measured DCR values were subtracted from the PCR values in all measurements.

In Fig.~\ref{BCR}(a), the bias current for the row SNSPDs is varied, while the column SNSPDs are not biased at all. The PCR curve for the \SI{3.4}{\micro\meter} narrow bandpass filter forms a plateau regime with only a small slope between \SI{23}{\micro\ampere} and \SI{31}{\micro\ampere}. Such a plateau is very desirable as it is usually interpreted as a regime of saturated internal detection efficiency of the SNSPDs. The slope in the plateau regime indicates the detection of unwanted blackbody radiation from the thermal light source assembly that is not filtered and reaches the array due to reflections inside the cryostat. The \SI{5.3}{\micro\meter} PCR curve has no pronounced plateau, but the curve has a distinct inflection point which indicates that the SNSPDs almost reach the regime of saturated internal detection efficiency. This is not true for the \SI{7.4}{\micro\meter} and the \SI{10}{\micro\meter} curve. This makes the detection efficiency of the array very prone to bias current fluctuations and indicates a reduced internal detection efficiency at these wavelengths, but the array can still be operated at these wavelengths. A very similar behavior can be observed in Fig.~\ref{BCR}(b), where only the column SNSPDs are biased and the count rates are measured at the column transmission line.

\begin{figure*}
    \centering
    \begin{subfigure}[b]{0.47\textwidth}
        \centering
        \includegraphics[width=1.00\textwidth]{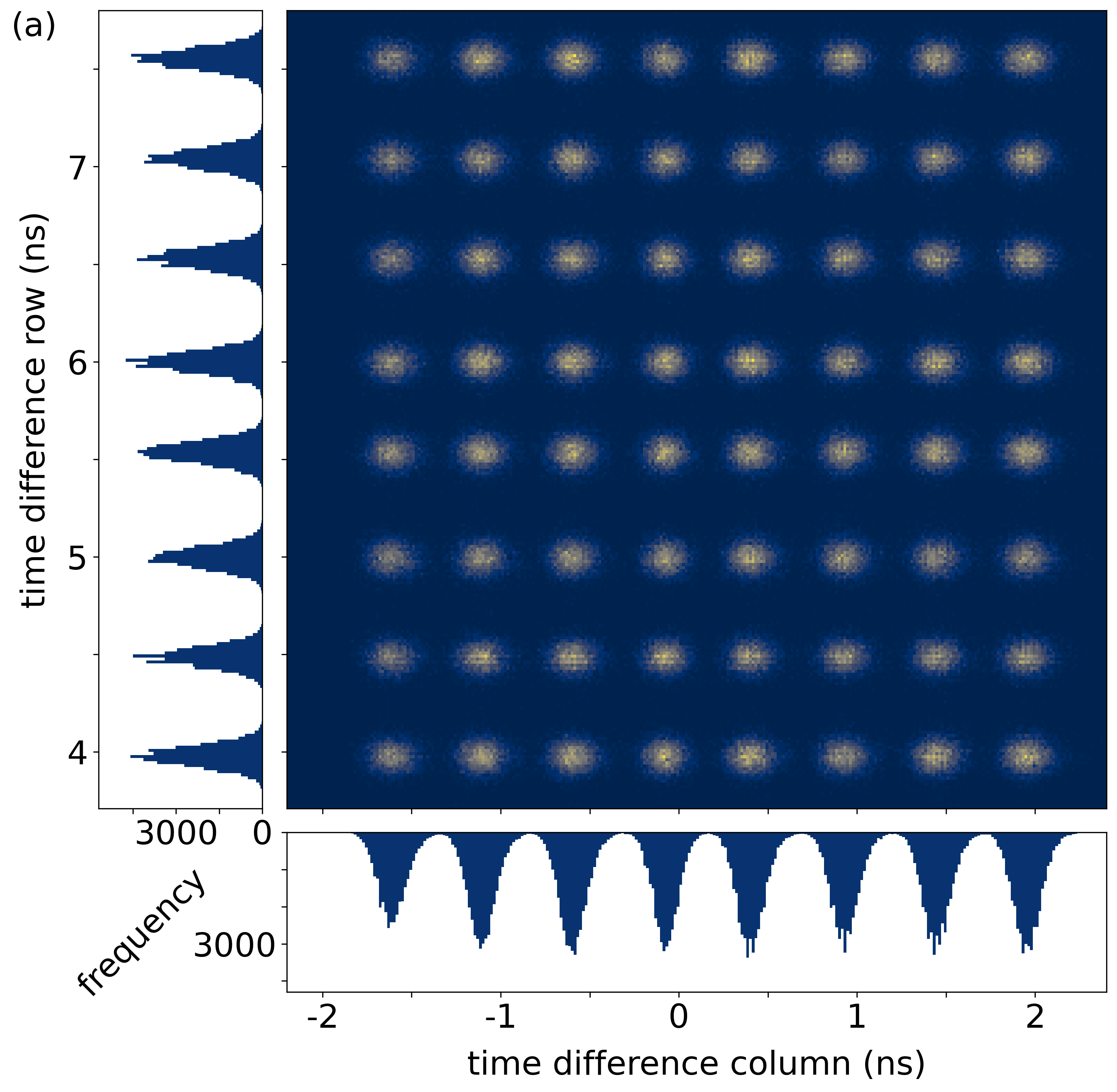}
        \label{histogram:3um}
        \vspace{-3mm}
    \end{subfigure}  
    \begin{subfigure}[b]{0.47\textwidth}
        \centering
        \includegraphics[width=1.00\textwidth]{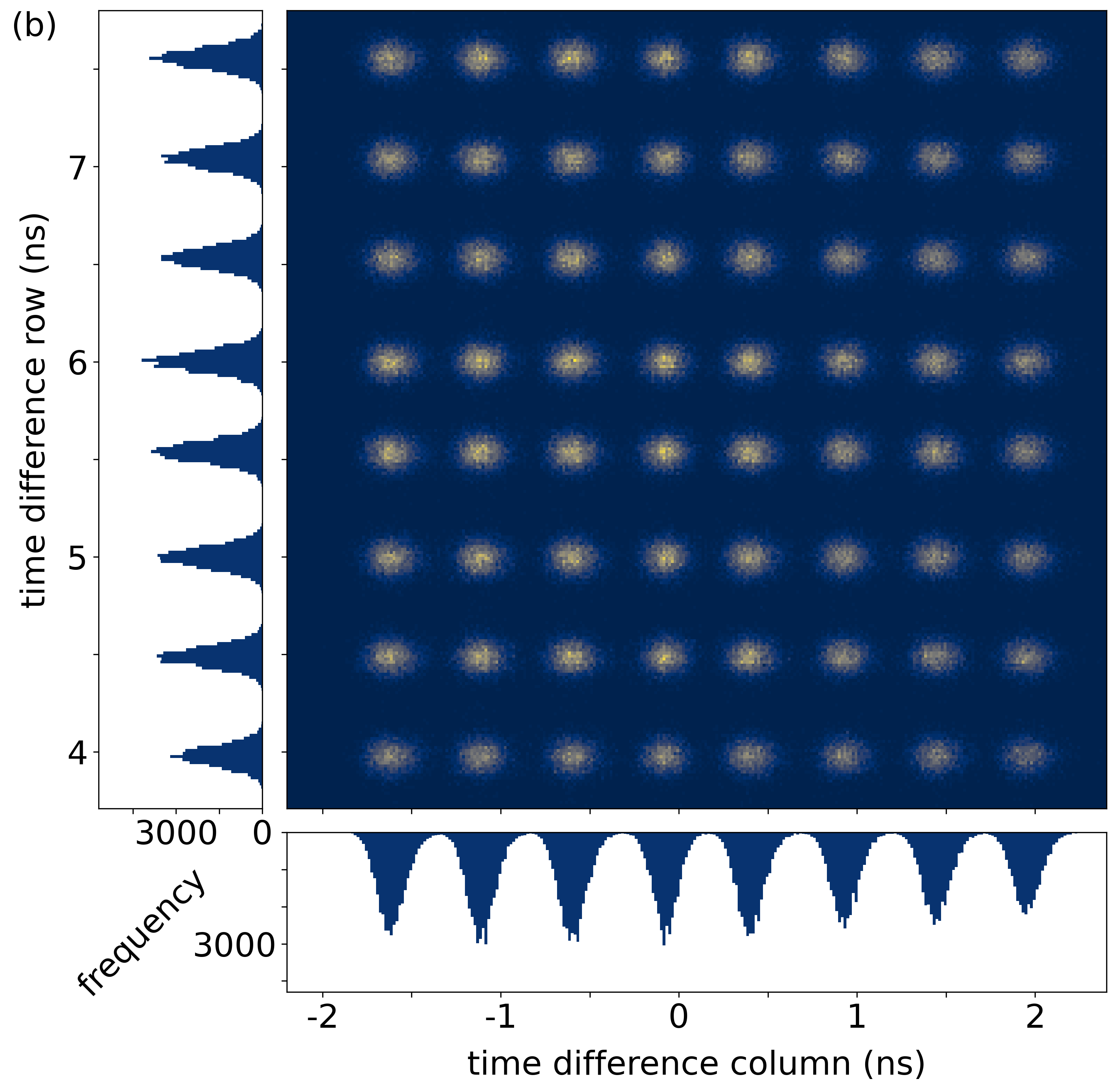}
        \label{histogram:5um}
        \vspace{-3mm}
    \end{subfigure} 
    \begin{subfigure}[b]{0.47\textwidth}
        \centering
        \includegraphics[width=1.00\textwidth]{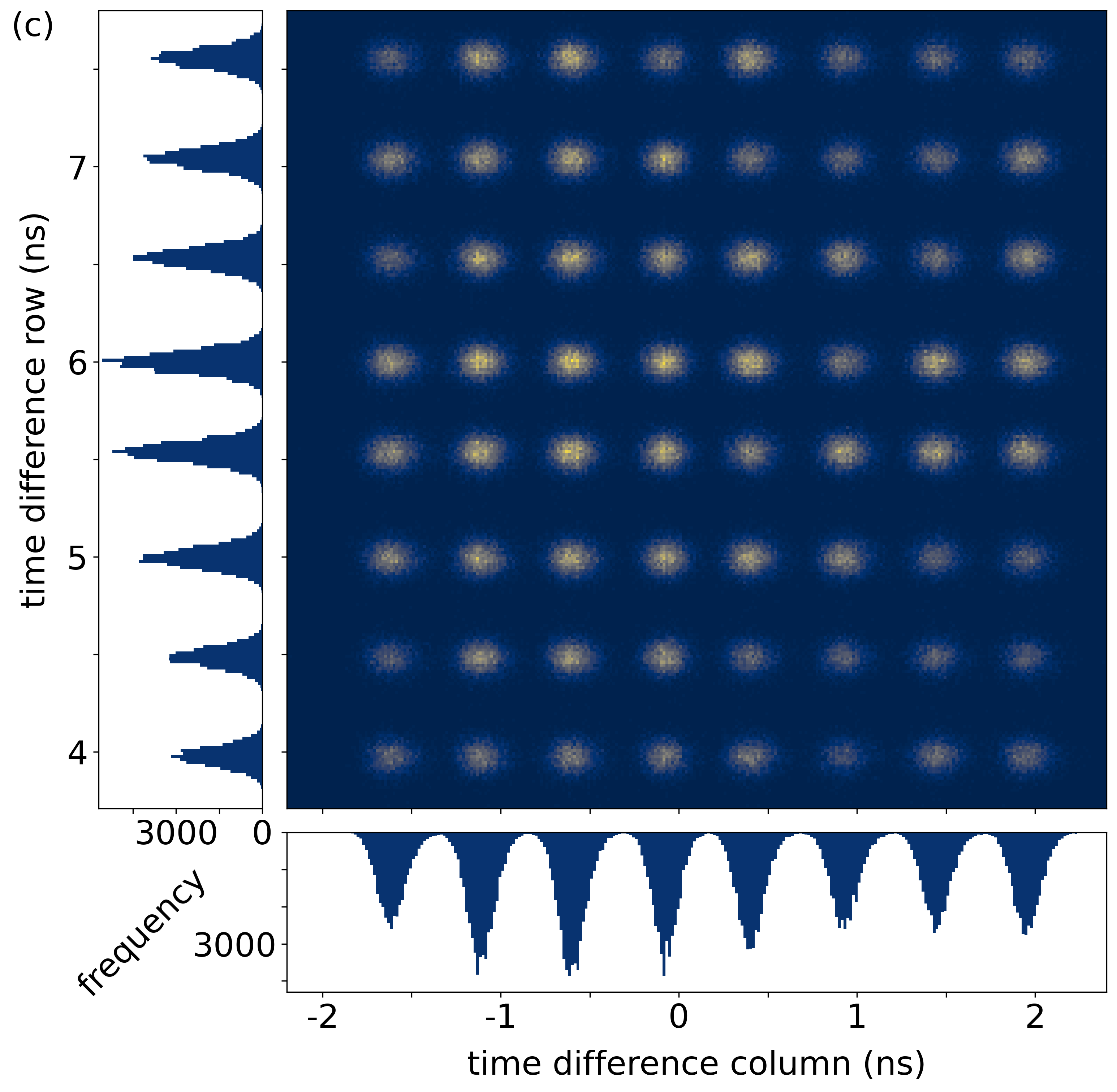}
        \label{histogram:7um}
        \vspace{-3mm}
    \end{subfigure} 
    \begin{subfigure}[b]{0.47\textwidth}
        \centering
        \includegraphics[width=1.00\textwidth]{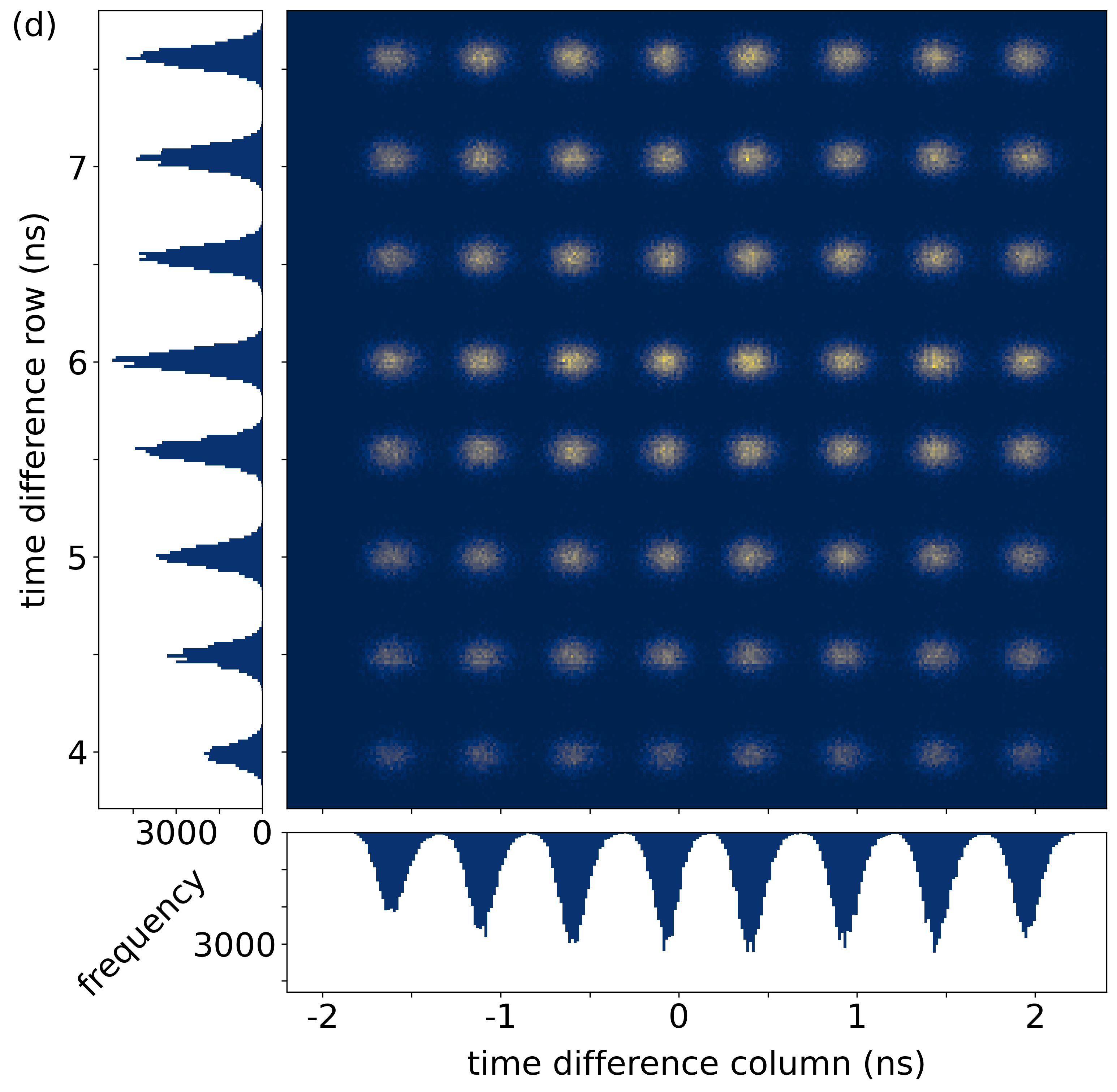}
        \label{histogram:10um}
        \vspace{-3mm}
    \end{subfigure} 
    \caption{Histograms of a \SI{1}{\second} time tagger measurement at the row and column transmission line for the thermal light source with narrow bandwidth filters at (a) \SI{3.4}{\micro\meter}, (b) \SI{5.3}{\micro\meter}, (c) \SI{7.4}{\micro\meter}, and (d) \SI{10}{\micro\meter}.}
    \label{histogram}
\end{figure*}

The measurement in Fig.~\ref{BCR}(c) is similar to Fig.~\ref{BCR}(a), but the column SNSPDs are biased with a fixed current of \SI{30}{\micro\ampere} in addition to the varying bias current in the row SNSPDs. Fig.~\ref{BCR}(d) can finally be compared to Fig.~\ref{BCR}(b) but biasing the row SNSPDs at \SI{30}{\micro\ampere} in addition to the varying bias current in the column SNSPDs. The PCR doubles for both of these measurement settings compared to the measurements in Fig.~\ref{BCR}(a) and (b), where only one of the two nanowire meanders in each pixel was biased (either row or column meander but not both). This behavior is expected as each detection event on either a row or the column nanowire leads also to a detection event on the co-wound nanowire due to thermal coupling. The fact that the PCR doubles is therefore a good indication that the thermal coupling works as intended. The elevated DCR in the measurement settings of Fig.~\ref{BCR}(c) and (d) is due to the fact that either the row or column SNSPDs are biased at \SI{30}{\micro\ampere} with a correspondingly high DCR that translates to a higher DCR due to thermal coupling between the co-wound SNSPDs.

Measurements with a time tagger are shown in Fig.~\ref{histogram}. Both the row and column SNSPDs are biased with a current of \SI{30}{\micro\ampere}, which corresponds to a bias current of \SI{3.75}{\micro\ampere} for each SNSPD, while both transmission lines are biased at \SI{2.8}{\micro\ampere}. The thermal light source was adjusted as described above to attain a PCR of \SI{100}{\kilo cps} for each wavelength. Both ends of both transmission lines are connected to the time tagger, which is configured to generate time stamps on a positive or negative voltage pulse, depending on the side of the transmission line. Each detection event will therefore generate four time stamps that can be correlated in post processing. The measurement time for each dataset for the four different narrow bandpass filters was \SI{1}{\second}. 

The dataset was used in post processing to calculate time differences between two correlated time stamps of the row and two of the column. This time difference is a measure for the position on the transmission line where the positive and negative voltage pulses were generated and therefore for the row or column that is responsible for the detection event. The calculated time differences for the rows and the columns are visualized as histograms in Fig.~\ref{histogram}. The mean full width at half maximum (FWHM) of all histogram peaks is \SI{122}{\pico\second} for the rows and \SI{167}{\pico\second} for the columns while all peaks are well separated by more than \SI{0.5}{\nano\second}. The two histograms for the rows and the columns are then combined to form a color-coded 2D-histogram that shows all 64 pixels. The array performs well for all tested mid-infrared wavelengths, but it can be seen that the distribution of detection events becomes inhomogeneous for longer wavelengths. This effect can be explained with the results presented in Fig.~\ref{BCR} where the SNSPDs did not reach saturated internal detection efficiency for longer wavelength photons, which made their detection efficiency prone to small nonuniformities in bias current across the array. Small resistance differences in the bias network of the SNSPD array could therefore affect the detection efficiency of whole rows and columns. This effect can significantly be reduced by optimizing the SNSPDs to achieve saturated internal detection efficiency at the target wavelength. The maximum count rate of the array will be limited by the maximum count rate of the transmission lines. The count rate of the transmission line remained linear as a function of the input photon flux up to \SI{2.5}{\mega cps}, with a \SI{3}{\decibel} compression point of \SI{6.5}{\mega cps}.\\

In summary, we have presented the design, fabrication, and measurement of a 64-pixel mid-infrared-optimized SNSPD array in this work. The performance of the array was tested at wavelengths up to \SI{10}{\micro\meter} and shows very promising results towards a large-scale mid-infrared camera. The fill factor of \SI{5}{\percent} could be further improved through reduction in size of the vias (interconnects between top and bottom layers) and wiring in the array active area. The membrane release could also be performed using a back side etch, eliminating the need for opening windows in the active area. Future work will also focus on optimization of system detection efficiency of single pixels, which will then need to be integrated into the array fabrication process. Simulations of both dielectric optical stacks and bow-tie antennas have shown that it should be possible to obtain greater than \SI{50}{\percent} detection efficiency over a \SI{1}{\micro\meter} bandwidth for a single pixel. A cryostat is also currently being constructed for calibrated efficiency measurements at mid-infrared wavelengths for both single pixels and arrays.

\vspace{5mm}

See the Supplementary Material for further details on the fabrication process of the SNSPD array.

\vspace{5mm}

We thank B.~G.~Oripov and A.~N.~McCaughan for helpful advice on design of the TCI architecture and fabrication, and B. Korzh for assistance with the thermal light source. We also thank M.~R.~Brann and B.~A.~Primavera for thoughtful comments on the manuscript. This research was funded by NIST (https://ror.org/05xpvk416). 

\section*{AUTHOR DECLARATIONS}
\vspace{-5mm}
\subsection*{Conflict of Interest} 
\vspace{-5mm}
The authors have no conflicts to disclose.

\subsection*{Author Contributions} 
\vspace{-5mm}
\textbf{Benedikt Hampel:} Conceptualization (equal); Data curation (lead); Formal analysis (lead); Investigation (lead); Methodology (equal); Software (lead); Validation (equal); Visualization (lead); Writing - original draft (lead); Writing - review \& editing (equal).
\textbf{Richard P. Mirin:} Funding acquisition (equal); Project administration (equal); Resources (supporting); Supervision (supporting);   Writing - review \& editing (equal).
\textbf{Sae Woo Nam:} Funding acquisition (equal); Project administration (equal); Resources (supporting); Supervision (supporting);  Writing - review \& editing (equal).
\textbf{Varun B. Verma:} Conceptualization (equal); Formal analysis (supporting); Funding acquisition (equal); Investigation (supporting); Methodology (equal); Project administration (equal); Resources (lead); Software (supporting); Supervision (lead); Validation (equal); Writing - original draft (supporting); Writing - review \& editing (equal).

\section*{Data Availability}
\vspace{-5mm}
The data that support the findings of this study are openly available in the NIST Science Data Portal at https://doi.org/XXXXXXXXXXXX.

\section*{References}
\bibliography{APL_MIRAR_TCI}

\end{document}


\title{A 64-pixel mid-infrared single-photon imager based on superconducting nanowire detectors} 

\author{Benedikt Hampel}
\affiliation{National Institute of Standards and Technology, Boulder, Colorado 80305, USA}
\affiliation{Department of Physics, University of Colorado, Boulder, Colorado 80309, USA}

\author{Richard P. Mirin}
\author{Sae Woo Nam}
\author{Varun B. Verma}
\affiliation{National Institute of Standards and Technology, Boulder, Colorado 80305, USA}


\maketitle 
\onecolumngrid

\vspace{-10mm}

\section*{Supplementary Material}

\section{Fabrication}
\twocolumngrid

Fabrication was performed on a 3-inch oxidized silicon (Si) wafer with a silicon dioxide (SiO\textsubscript{2}) thickness of \SI{150}{\nano\meter}. A \SI{5}{\nano\meter}-thick layer of tungsten silicide (W\textsubscript{0.5}Si\textsubscript{0.5}) was deposited by dc magnetron co-sputtering from separate tungsten (W) and Si targets at room temperature, followed by a \SI{2}{\nano\meter}-thick layer of amorphous silicon. The amorphous silicon layer protects and prevents oxidation of the WSi layer during subsequent processing steps. Electron beam lithography was then performed using polymethyl methacrylate (PMMA) resist to define the co-wound SNSPD pixels and the heater and transmission line bus constrictions. The patterns were transferred into the underlying WSi superconductor layer using reactive ion etching (RIE) in an sulfur hexafluoride (SF\textsubscript{6}) plasma. The SNSPDs consist of nanowires which are \SI{180}{\nano\meter} wide. The co-wound row nanowire and column nanowire in each pixel are separated by a \SI{100}{\nano\meter}-gap. Each pixel covers an area of \qtyproduct{10 x 5}{\micro\meter}. Adjacent pixels are \SI{30}{\micro\meter} apart, yielding a fill factor of \SI{5}{\percent}. The superconducting constriction corresponding to \textit{R}\textsubscript{heater} is \SI{60}{\nano\meter}-wide. This constriction must be significantly smaller than the nanowire width used to form the SNSPD to ensure that it switches to the normal state when the SNSPD bias current is diverted to it. \textit{R}\textsubscript{heater} is separated from the superconducting constriction corresponding to \textit{R}\textsubscript{bus} by a \SI{100}{\nano\meter}-wide gap. The width of the constriction on the bus is \SI{180}{\nano\meter}. 

Larger areas of WSi such as the transmission line bus, inductors, and wiring between components were patterned using optical lithography and also etched by RIE in an SF\textsubscript{6} plasma. The inductor \textit{L}\textsubscript{bias} consists of a \SI{2}{\micro\meter}-wide meandering wire with a total length of \SI{36}{\milli\meter}. With a kinetic inductance for WSi of \SI{550}{\pico\henry\per\sq}, this yields an inductance of \textit{L}\textsubscript{bias}~=~\SI{10}{\micro\henry}. Resistors were patterned by optical lithography and liftoff of a metal stack consisting of \SI{2}{\nano\meter} titanium (Ti) and \SI{25}{\nano\meter} palladium gold (PdAu). A 30 second argon (Ar) clean was performed before deposition of the resistor metals to remove residual photoresist and remove the amorphous silicon cap layer covering the WSi in the regions where the WSi makes contact to the resistors. The resistor geometries were chosen to provide for the following resistance values: \textit{R}\textsubscript{bias}~=~\SI{130}{\ohm}, \textit{R}\textsubscript{shunt}~=~\SI{25}{\ohm}, and \textit{R}\textsubscript{tc}~=~\SI{25}{\ohm}. After deposition and liftoff of the resistor layer, another photolithography and metal liftoff step was performed to define metal pads serving as interconnects between the WSi in the bottom layer, and the top wiring layers and ground plane which will later be deposited on top of a layer of \SI{50}{\nano\meter} of SiO\textsubscript{2} which covers the WSi layer. These pads also serve as an etch stop for the vias when etching through the top SiO\textsubscript{2} layer. The metal stack for the pads consists of \SI{2}{\nano\meter} Ti, \SI{50}{\nano\meter} niobium (Nb), \SI{10}{\nano\meter} PdAu, and \SI{2}{\nano\meter} Ti. The wafer is then cleaned in an oxygen (O\textsubscript{2}) plasma for 1 minute at \SI{60}{\watt}. Next, a \SI{50}{\nano\meter}-thick layer of SiO\textsubscript{2} is deposited by radio frequency (rf) sputtering. The thickness of this layer is important as it sets the capacitance between the superconducting microstrip transmission line and the ground plane, which along with the kinetic inductance of the WSi strip forming the transmission line determines the propagation velocity of the electrical pulses along the transmission line. With a kinetic inductance for WSi of \SI{550}{\pico\henry\per\sq}, the propagation velocity is only \SI{0.5}{\percent} the speed of light in free space. This slow propagation velocity combined with the small jitter of SNSPDs allows pulses from adjacent rows and columns to be easily discriminated.  

After deposition of the SiO\textsubscript{2}, vias were defined by photolithography and the SiO\textsubscript{2} was etched down to the underlying metal pads using a reactive ion etch in trifluoromethane (CHF\textsubscript{3}) and O\textsubscript{2}. This was followed by another photolithography and liftoff step to define the bond pads, ground plane, and interconnects for the column bias current in the SNSPD array as shown in Fig.~1(d). This layer consists of \SI{100}{\nano\meter} of Nb. The last step of the process is the suspension of the SiO\textsubscript{2} membranes underneath the SNSPD pixels and the coupling region between the heater constriction and bus constriction. Two rectangular windows are patterned using optical lithography above and below each SNSPD pixel in the array shown in Fig.1(d), as well as above and below the coupling regions between heater and bus constrictions shown in Fig.~1(a). An inductively coupled plasma reactive ion etch using tetrafluoromethane (CF\textsubscript{4}), Ar, and O\textsubscript{2} is used to etch through the \SI{200}{\nano\meter} of SiO\textsubscript{2} down to the silicon substrate. The photoresist is not removed after this etch because the same pattern will be used in the undercut of the silicon substrate. The substrate is then diced into \SI{1}{\centi\meter} chips, and a second reactive ion etch is performed for 4 minutes in an SF\textsubscript{6} plasma at a high pressure of \SI{200}{\milli\torr} and low rf power to provide for an isotropic etch of the silicon substrate underneath the SiO\textsubscript{2}. The photoresist is removed in an O\textsubscript{2} plasma at \SI{100}{\watt} for 8 minutes.